\documentclass[12pt,journal,draftclsnofoot,a4paper,oneside,onecolumn]{IEEEtran}

\usepackage{amsfonts}
\usepackage{mathrsfs}
\usepackage{cite}
\usepackage{graphicx}
\usepackage{amsmath}
\usepackage{amssymb}
\usepackage{stfloats}
\usepackage{subeqnarray}
\usepackage{cases}
\usepackage{indentfirst}
\usepackage{bm}
\usepackage{setspace}

\begin{document}
%
\title{\huge{Truthful Mechanisms for Secure Communication in Wireless Cooperative System}}
%
%

\author{\IEEEauthorblockN{\normalsize{Jun Deng}, Rongqing Zhang,  \IEEEmembership{\normalsize{Student Member, IEEE}}, Lingyang Song,  \IEEEmembership{\normalsize{Senior Member, IEEE}}, \\
Zhu Han,  \IEEEmembership{\normalsize{Senior Member, IEEE}} and Bingli
Jiao,  \IEEEmembership{\normalsize{Senior Member, IEEE}}}\\

\thanks{Part of this work has been published in INFOCOM WKSHPS 2011 as given in \cite{liushuhang}.}


\thanks{Jun Deng, Rongqing Zhang, Lingyang Song and Bingli Jiao are with
State Key Laboratory of Advanced Optical Communication Systems and
Networks, School of Electronics Engineering and Computer Science,
Peking University, Beijing, China (email: \{dengjun,
rongqing.zhang, lingyang.song, jiaobl\}@pku.edu.cn).

Zhu Han is with Electrical and Computer Engineering Department,
 University of Houston, Houston, USA (email: zhan@uh.edu).}

}

\maketitle

\begin{abstract}

To ensure security in data transmission is one of the most important
issues for wireless relay networks, and physical layer security is an attractive
alternative solution to address this issue. In this paper, we consider a
cooperative network, consisting of one source node, one destination
node, one eavesdropper node, and a number of relay nodes.
Specifically, the source may select several relays to help
forward the signal to the corresponding destination to achieve the
best security performance. However, the relays may have the
incentive not to report their true private channel information in
order to get more chances to be selected and gain more payoff from
the source. We propose a Vickey-Clark-Grove (VCG) based mechanism and an
Arrow-d'Aspremont-Gerard-Varet (AGV) based mechanism into the investigated relay
network to solve this cheating problem.
In these two different mechanisms, we design different ``transfer
payment'' functions to the payoff of each selected relay and
prove that each relay gets its maximum (expected) payoff when it truthfully reveals
its private channel information to the source. And then, an optimal
secrecy rate of the network can be achieved. After discussing and
comparing the VCG and AGV mechanisms, we prove that the AGV mechanism can
achieve all of the basic qualifications (incentive compatibility,
individual rationality and budget balance) for our system. Moreover,
we discuss the optimal quantity of relays that the source node
should select. Simulation results verify efficiency and fairness
of the VCG and AGV mechanisms, and consolidate these conclusions.

\end{abstract}


\newpage

\section{Introduction}
\IEEEPARstart{S}{ecurity} is one of the most important issues
in wireless communications due to the broadcast nature of wireless radio
channels. In recent years, besides the traditional cryptographic
mechanisms, information-theoretic-based physical layer security has
been developing fast. The concept of ``wiretap channel'' was first
introduced by Wyner \cite{AW-1975}, who showed that perfect secrecy
of transmitted data from the source to the legitimate receiver is
achievable in degraded broadcast channels. In follow-up work,
Leung-Yan-Cheong and Hellman further determined the secrecy capacity
in the Gaussian wire-tap channel \cite{CH-1978}. Later, Csiszar and
Komer extended Wyner's work to non-degraded broadcast channels and
found an expression of secrecy capacity \cite{CK-1978}.

When considering a wireless relay network, realization of
secrecy capacity is much more complicated. In \cite{ASCP-2009}, the
authors studied the secrecy capacity of a relay channel with
orthogonal components in the presence of a passive eavesdropper
node. In \cite{DHPP-2010, LPW-2011}, the authors demonstrated that cooperation
among relay nodes can dramatically improve the physical layer
security in a given wireless relay network, And in \cite{RQ-2010,JC-2011,HS-2011},
the authors investigated the physical layer security with friendly jammer
in the relay networks. In the related work mentioned above, the channel state
information (CSI) is assumed to be known at both the transmitter and
the receiver. All these schemes are assumed under true channel information reported by relay nodes, and the optimal
solutions in these works will not hold anymore if the fake channel information is reported.
However, in practice, the relay
node always measures its own channel gains and distributes the
information to others through a control channel. There is no
guarantee that it reveals its private information honestly. Hence,
the most critical problem is how to select efficient relay nodes to
optimize the total secrecy rate in the network, while some selfish
relays may report false information to the source to increase their
own utilities. In \cite{RMS-2005}, the
reputation methods are designed to achieve this goal. However, all
these methods require a delicate and complex ``detection scheme'' to
monitor and capture the liar nodes.  It needs a
lot of signal consumption like an independent entity called ``Trust Manager" to runs on each
node. Besides, it also demands large amount of data
like ``REP\_MESSAGE",``REP\_VAL" to record the intermediate variable in the process. Moreover,
this method requires long time of observation because of low speed of convergence. It might be impractical to use these
reputation based scheme in cooperative relay network.

In recent years, the game theory is widely applied into wireless and
communication networks to solve resource allocation
problems\cite{AS-2001,YFan-2011,HDA-2010,JANB-2011,ZDWTA-2011,RQ-2011}. In the area of
mechanism design, a field in the game theory studying solution concepts
for a class of private information games, a game designer is
interested in the game's outcome and wants to motivate the players
to disclose their private information by designing the payoff
structure \cite{AMJ-1995,NA-1999,YDRH-2009}. For example, the
well-known VCG (Vickrey-Clarke-Groves) mechanism
\cite{WC-1961,EC-1971,TG-1973} is a dominant-strategy mechanism,
which can achieve ex-post incentive compatibility (truth-telling
is a dominant strategy for every player in the game). However, it
cannot implement the budget balance of the game
\cite{HS-2001,PB-2007}, which costs extra payment from the
players and decrease their payoffs. Thus, it cannot be properly used
in the relay network we focus on. Compared with the VCG mechanism, the AGV
(Arrow-d'Aspremont-Gerard-Varet) mechanism \cite{KJA-1979,CL-1979}
can also solve the truth-telling problem. It is an incentive
efficient mechanism that can maximize the expected total payoff of
all the players in the game. Additionally it achieves the budget balance
under a weaker participation requirement \cite{FT-1993,Vk-2002}.

In this paper, we mainly focus on a relay network, in which all the
channels are orthogonal and each relay's private channel information
is unknown by others. Under these conditions, we apply the ideas of
the VCG and AGV mechanism and prove that the transfer function can meet the
basic requirements of the wireless relay networks and help achieve
the truth-telling target. We find and prove that the unique Bayesian
Nash Equilibrium \cite{FT-1993} is achieved when all the relays in
the network reveal the truth. The incentive to report false information
will lead to a loss in each relay node's own (expected) payment.
In other words, the competing relay nodes are enforced to obey the selection criterion and cooperate with each
other honestly. Furthermore, there is no extra cost paid in the
system when applying the AGV mechanism while the VCG mechanism can not. Since the AGV mechanism is
budget balanced, which means the total transfer payment of all relay
nodes equals zero. Simulation results show that the relay nodes can maximize
their utilities when they all report their true channel information.
Any cheating to the source leads to certain loss in the total secrecy
rate as well as the payoff of relays themselves. We also observe
that the optimal choice for the system is based on the relays'
channel information, but in a majority of cases selecting only
one relay node for transmitting data can attain the largest secrecy
rate of the system. In addition, we prove with simulations that the
best strategy for each relay node under this payoff structure is to
improve its own channel condition to enlarge its secrecy rate and
always report the truth to the source.

The remainder of this paper is organized as follows. In Section
\uppercase\expandafter{\romannumeral2}, the system model for a
relay network is presented. In Section
\uppercase\expandafter{\romannumeral3}, we elaborate on the basic
definition and qualifications of the mechanism design, and discuss
the VCG mechanism and AGV mechanism. In Section
\uppercase\expandafter{\romannumeral4}, we demonstrate the mechanism
solutions to enforce relays reveal the true private information, and
analyze these mechanism solutions. Simulation results are shown in
Section \uppercase\expandafter{\romannumeral5}, and the conclusions
are drawn in Section~\uppercase\expandafter{\romannumeral6}.

\section{System Model}%
\begin{figure}
\centering
\includegraphics[width=4.8in]{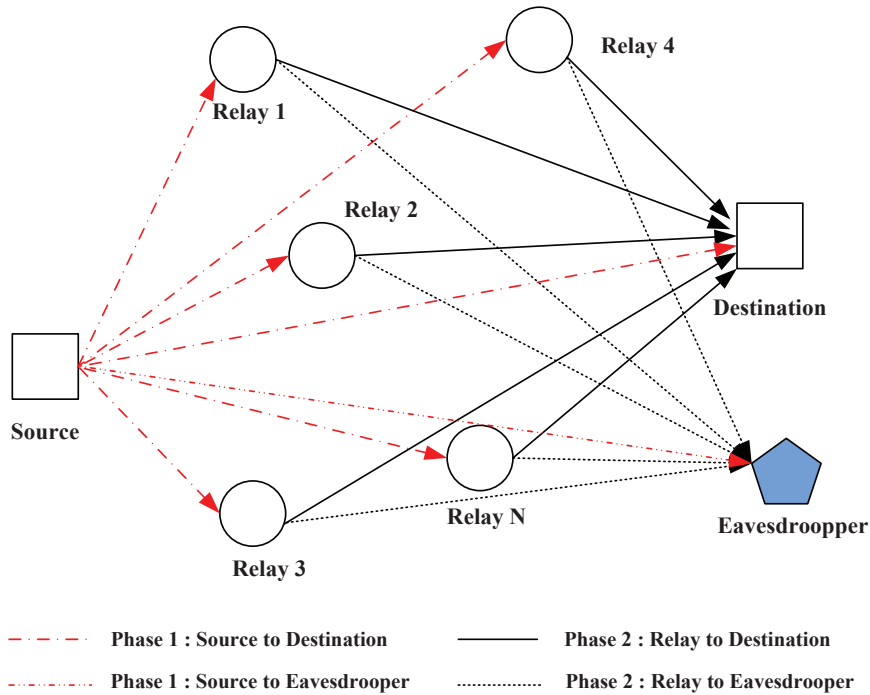}
\caption{System model for relay network with eavesdropper.}
\label{fig_1}
\end{figure}

Considering a general cooperative network shown in Fig.~\ref{fig_1}.
It consists of one source node, one destination node, one
eavesdropper node, and $I$ relay nodes, which are denoted by $S$,
$D$, $E$, and $R_i$, $i = 1,2,\ldots,I$, respectively. This
cooperative network is conducted in two phases. In phase 1, the
source node broadcasts a signal $x$ to the destination node and all
the relay nodes, where only $N$ $(N \le I)$ nodes can decode this
signal correctly due to their different geographical conditions. In
phase 2, the source node decides which nodes of those $N$ relays to forword
information to the destination node. The destination node combines
messages from the source and relays, according to the reported
channel gains of both relay-destination and relay-eavesdropper
links. During the whole process, eavesdropper node wiretaps the
messages from the source node and the relay nodes. We assume the orthogonal
channel having the same bandwidth $W$. The source node hopes to
gain the highest secrecy rate by properly selecting some efficient
relay nodes based on their reported channel information. We denote
the number of selected relay nodes by $K$ $(K \le N)$ and the set of
$K$ relay nodes by $\mathcal{K}$.

In the first phase, the received signal $y_{s,d}$, $y_{s,r_i}$, and
$y_{s,e}$ at destination node $D$, relays $R_i$, and
eavesdropper $E$, respectively, can be expressed as
\begin{align}
y_{s,d} = \sqrt{P_s}h_{s,d}x+n_{s,d},
\end{align}
and
\begin{align}
y_{s,r_i} = \sqrt{P_s}h_{s,r_i}x+n_{s,r_i},
\end{align}
and
\begin{align}
y_{s,e} = \sqrt{P_s}h_{s,e}x+n_{s,e},
\end{align}
where $P_s$ represents the transmit power to the destination node
from the source node, $x$ is the unit-energy information symbol
transmitted by the source in phase 1, $h_{s, d}$, $h_{s,r_i}$, and
$h_{s,e}$ are the channel gains from $S$ to $D$, $R_i$ and $E$
respectively. $n_{s,d}$, $n_{s,r_i}$ and $n_{s,e}$ represent the noise
at destination node, relay nodes and eavesdropper node.

In the second phase, the received signal from
the $i$-th relay node $(R_i\in\mathcal{K})$ to the destination node and
eavesdropper node can be expressed as
\begin{align} \label{eq1}
y_{r_i ,d}  = \sqrt {P_{r_i } } h_{r_i ,d} x + n_{r_i ,d},
\end{align}
and
\begin{align} \label{eq2}
y_{r_i ,e}  = \sqrt {P_{r_i } } h_{r_i ,e} x + n_{r_i ,e},
\end{align}
respectively, where $P_{r_i}$ denotes the transmit power of relay node $R_i$ under
the power constraint $P_{r_i} \le P_{max}$, $h_{r_i,d}$ is the
channel gain between $R_i$ and $D$, and $h_{r_i,e}$ is the channel
gain between $R_i$ and $E$. We assume that channel gain
contains both the path loss and the Rayleigh fading
factor. Without loss of generality, we also assume that all the
links have the same noise power which is denoted by ${\sigma ^2}$.
The decode-and-forward~(DF) protocol is used for relaying.

The direct transmission signal-to-noise-ratios (SNR) at the
destination node and eavesdropper from the source are
$\mbox{SNR}_{s,d} = \frac{P_sh_{s,d}^2}{\sigma^2},$ and
$\mbox{SNR}_{s,e} = \frac{P_s h_{s,e}^2} {\sigma^2},$
respectively. The SNR at the destination node and
eavesdropper node from relays are
$\mbox{SNR}_{r_i,d}  = \frac{{P_{r_i } h_{r_i ,d}^2 }}{{\sigma ^2 }},$ and
$\mbox{SNR}_{r_i,e}  = \frac{{P_{r_i } h_{r_i ,e}^2 }}{{\sigma ^2 }}.$
Therefore, the channel rate for relay $R_i$ to destination $D$ is
\begin{align} \label{eq5}
C_{i,d}  = W\log _2 \left(1 + \mbox{SNR}_{r_i,d} \right).
\end{align}
Similarly, the channel rate for relay $R_i$ to eavesdropper $E$ is
\begin{align} \label{eq6}
C_{i,e}  = W\log _2 (1 + \mbox{SNR}_{r_i,e} ).
\end{align}
Then, the secrecy rate achieved by $R_i$ can be defined
as~\cite{BR-2006}
\begin{align} \label{eq7}
{C_{i,s}}  = {\left( {{C_{i,d}} - {C_{i,e}}} \right)^ + }
    = {\left[ {W{{\log }_2}\left( {\frac{{1 + \frac{{{P_{{r_i}}}h_{{r_i},d}^2}}{{{\sigma ^2}}}}}{{1 + \frac{{{P_{{r_i}}}h_{{r_i},e}^2}}{{{\sigma ^2}}}}}} \right)} \right]^ +
    },
\end{align}
where $(x)^+ = \max \{x,0\}$.

Besides, the secrecy rate assuming maximal ratio combining (MRC) at
the destination and the eavesdropper can be written as
\begin{align}
C_{d,sys} =W\log_2\left(1+\mbox{SNR}_{s,d} + \sum\limits_{i \in
\mathcal{K}} {\mbox{SNR}_{r_i,d}}\right)
\end{align}
and
\begin{align}
C_{e,sys} = W\log_2\left(1+\mbox{SNR}_{s,e}+ \sum\limits_{i \in
\mathcal {K}} {\mbox{SNR}_{r_i, e}}\right),
\end{align}
respectively, such that the total secrecy rate attained by the system is
\begin{align}\nonumber \label{eq15}
 {C_{s,sys}} &= \left({C_{d,sys}} - {C_{e,sys}}\right)^+ \\\nonumber
  &= W{\log _2} \left(1 + \mbox{SNR}_{s,d} + \sum\limits_{i \in \mathcal {K}} {\mbox{SNR}_{r_i,d}} \right) \\
  &- W{\log _2}\left(1 + \mbox{SNR}_{s,e} + \sum\limits_{i \in \mathcal {K}} {\mbox{SNR}_{{r_i},e}}
  \right) \\
  &= W{\log _2}\left(\frac{{1 + \mbox{SNR}_{s,d} + \sum\limits_{i \in \mathcal {K}} {\mbox{SNR}_{r_i,d}} }}{1 + \mbox{SNR}_{s,e} + \sum\limits_{i \in \mathcal {K}} {\mbox{SNR}_{{r_i},e}}}\right).\nonumber
\end{align}

\section{Mechanism Design}%
In this paper, we use mechanism design as the framework to create an efficient way to prevent relay nodes
from cheating in the process of selection. This section provides an overview
of essential concepts in mechanism design, the VCG mechanism and AGV mechanism.

\subsection{Basic Definitions and Qualifications}
Consider a public system consisting of $I$ agents, ${1, 2, \ldots
I}$. Each agent $i \in \{ 1,2 \ldots I\}$ has its private
information ${\theta _i} \in {\Theta _i}$, which is known by itself
only. A social choice function $F$ is defined as
\begin{align}\nonumber
F:{\Theta _1} \times {\Theta _2} \times  \ldots  \times {\Theta _I}
\to O,
\end{align}
where $O$ stands for a set of possible outcomes.

A mechanism $\mathcal {M}$ is represented by the tuple $(F, t_1,
\ldots, t_I)$, where $t_i$ is the transfer payment of agent $i$ when
the social choice is $F$. The utility of agent $i$:
$v_i\left[F\left({{\hat \theta }_i},{{\hat \theta }_{ - i}} \right),
\theta_i\right]$ depends on the outcome $o=F\left({{\hat \theta
}_i},{{\hat \theta }_{ - i}} \right)$ and the true information of
agent $i$: $\theta_i$, where ${\hat \theta}_i$ denotes the reported
information of agent $i$, as opposed to ${\theta_i}$. Similarly,
${\hat \theta}_{-i}=\{{\hat\theta_1, \ldots, \hat\theta_{i-1}, \hat\theta_{i+1}, \ldots,
\hat\theta_I}\}$ is the reported information of all other agents. So the
total payoff or welfare of agent $i$ can be written as the following
function:
\begin{align}
{u_i}\left({{\hat \theta }_i},{{\hat \theta }_{ - i}},{\theta
_i}\right) = v_i\left[F\left({{\hat \theta }_i},{{\hat \theta }_{ -
i}}\right),{\theta _i}\right] + {t_i}\left({{\hat \theta }_i},{{\hat
\theta }_{ - i}}\right).
\end{align}
The objective of mechanism $\mathcal{M}$ is to choose a desirable
set of transfer payments $t_i$. Thus, each agent in the mechanism
will achieve its maximum payoffs. In the following, we define some
properties for the mechanism.

\emph{Definition 1}: A mechanism is
\emph{incentive compatible (IC)} if the truth-telling is the best strategy for
the agents: ${\hat \theta} _i = \theta_i$, which means that agents have
no incentives to reveal false information. The dominant-strategy IC
is defined as
\begin{align}
{u_i}\left({\theta _i},{{\hat \theta }_{ - i}},{\theta _i}\right)
\ge {u_i}\left({{\hat \theta }_i},{{\hat \theta }_{ - i}},{\theta
_i}\right),\;\;\;\forall {{\hat \theta }_i},{\theta _i} \in
{\Theta _i},{{\hat \theta }_{ - i}} \in {\Theta _{ - i}}.
\end{align}

\emph{Definition 2}: In an \emph{individual
rational (IR)} mechanism, rational agents are expected to gain a higher
utility from actively participating in the mechanism than from
avoiding it. Especially, in the dominant strategy IR can be
expressed as
\begin{align}
{u_i}\left({\hat\theta _i},{\hat\theta _{ - i}},{\theta _i}\right) \ge
0,\;\forall {\theta _i} \in {\Theta _i}.
\end{align}
A mechanism that is both incentive-compatible and individual
rational is said to be \emph{strategy-proof}.

\emph{Definition 3}: In a \emph{budget balanced (BB)}
mechanism, the sum of all agents transfer payments is zero, which
implies that there is no transfer payment paid from the mechanism
designer to the agents or the other way around. The BB is defined
as
\begin{align}
\sum\limits_{i = 1}^I {{t_i}\left({{\hat \theta }_i},{{\hat \theta
}_{ - i}}\right)}  = 0,\;\forall {\theta _i} \in {\Theta
_i}.
\end{align}

\subsection{VCG Mechanism}%
Groves introduced a group of mechanisms which satisfy IC and IR.
The Groves mechanisms are characterized by the following transfer
payment function:
\begin{align}
{t_i}\left({{\hat \theta }_i},{{\hat \theta }_{ - i}}\right) =
\sum\limits_{j \ne i}^I {{v_j}\left[F\left({{\hat \theta }_j},{{\hat
\theta }_{ - j}}\right),{{\hat \theta }_j}\right]}  - {\tau
_i}\left({{\hat \theta }_{ - i}}\right),
\end{align}
where $\tau_i(.)$ can be any function of ${\hat \theta}_i$. The VCG
mechanism is an important special case of
the Groves mechanisms for which
\begin{align}
{\tau _i}\left({{\hat \theta }_{ - i}}\right) = \sum\limits_{j \ne
i}^I {{v_j}\left[{F^*}\left({{\hat \theta }_j},{{\hat \theta }_{ -
j}}\right),{{\hat \theta }_j}\right]},
\end{align}
where ${{F^*}\left({{\hat \theta }_j},{{\hat \theta }_{ -
j}}\right)}$ is the outcome of the mechanism when agent $i$
withdraws from the mechanism. Thus, in the VCG mechanism agent $i$
could attain payoff as
\begin{align}\nonumber
 {u_i}\left({{\hat \theta }_i},{{\hat \theta }_{ - i}},{\theta _i}\right) &= {v_i}\hspace{-0.1cm}\left[F\left({{\hat \theta }_i},{{\hat \theta }_{ - i}}\right),{\theta _i}\right]\hspace{-0.1cm} + {t_i}\left({{\hat \theta }_i},{{\hat \theta }_{ -
 i}}\right)\\\nonumber
  &= {v_i}\hspace{-0.1cm}\left[F\left({{\hat \theta }_i},{{\hat \theta }_{ - i}}\right),{\theta _i}\right] \hspace{-0.1cm} +\hspace{-0.1cm} \sum\limits_{j \ne i}^I {{v_j}\hspace{-0.1cm}\left[F\left({{\hat \theta }_j},{{\hat \theta }_{ - j}}\right),{{\hat \theta }_j}\right]}  \\
  &- \sum\limits_{j \ne i}^I {{v_j}\hspace{-0.1cm}\left[{F^*}\left({{\hat \theta }_j},{{\hat \theta }_{ - j}}\right),{{\hat \theta
  }_j}\right]}.
\end{align}

As will be proved in the next section, the VCG mechanism can satisfy both the incentive compatibility and
individual rationality of each agent. However, the VCG mechanism is not budget-balanced and
requires a third party agent to mediate between mechanism designer
and agents.

\subsection{AGV Mechanism}%
The AGV mechanism, an extension of
the Groves mechanism, is possible to
achieve IC, IR and BB. It is an ``expected form" of the Groves
mechanism and its transfer payment function is defined as
\begin{align}
{t_i}\left({{\hat \theta }_i},{{\hat \theta }_{ - i}}\right) =
{E_{{\theta _{ - i}}}} \left\{\sum\limits_{j \ne i}^I
{v_j}\left[F\left({{\hat \theta }_j},{{\hat \theta }_{ -
j}}\right),{{\hat \theta }_j}\right] \right\}  - {\tau
_i}\left({{\hat \theta }_{ - i}}\right).
\end{align}

The first term of $t_i$ is the expected total utility of agents $j
\ne i$ when agent $i$ reports its information ${\hat \theta}_i$ with
the assumption that other agents report the truth. It is the
function of agent $i$'s report information only, exclusive of the
actual strategies of agents $j \ne i$, which making the AGV mechanism
different from the VCG mechanism.

In the AGV mechanism it is possible to design the $\tau_i(.)$ to satisfy
BB. Let
\begin{align}
{\Phi _i}\left({{\hat \theta }_i}\right) = {E_{{\theta _{ -
i}}}}\left\{\sum\limits_{j \ne i}^I {v_j}\left[F\left({{\hat \theta
}_j},{{\hat \theta }_{ - j}}\right),{{\hat \theta }_j}\right]\right\}
\end{align}
and
\begin{align}
{\tau _i}\left({{\hat \theta }_{ - i}}\right) = \frac{{ -
\sum\limits_{j \ne i}^I {{\Phi _j}\left({{\hat \theta }_j}\right)}
}}{{I - 1}},
\end{align}
then budget balance can be achieved because each agent also pays an
equal share of total transfer payments distributed to the other
agents, none of which depends on its own report information. We will
prove this property in the following section.

\section{Mechanism Solutions}%

In this section, we first describe how mechanism design is applied
in the relay system and prove that there is no equilibrium achieved
if no transfer payment is introduced to relay nodes. Then, we show the
practical mapping from the utility, transfer payment, and payoff of
the VCG mechanism and AGV mechanism to the wireless cooperative
network. Finally, we will compare and analyze the difference
between two mechanisms.
\subsection{Mechanism Implementation}
In the network, each relay node reports its own channel information
$(h_{r_i,d}, h_{r_i,e})$ to the source node which can be seen as
different agents report their own private information to mechanism
designer. Assume $\left\{\left({{\tilde h}_{{r_1},d}},{{\tilde
h}_{{r_1},e}}\right),\left({{\tilde h}_{{r_2},d}},{{\tilde
h}_{{r_2},e}}\right), \ldots ,\left({{\tilde h}_{{r_K},d}},{{\tilde
h}_{{r_K},e}}\right)\right\}$ is a realization of channel gains at one time slot, and
relay nodes report their information $\left\{\left(\hat h_{r_1,d},\hat h_{r_1,e}\right)\right.,$ $\left(\hat
h_{r_2,d},\hat h_{r_2,e}\right),$ $\ldots,$ $\left.\left(\hat h_{r_K,d},\hat
h_{r_K,e}\right)\right\} $
to the source node. Though the information may not be true,
the source node will still select relay nodes based on them. Define $R_i$'s
private channel information as $\tilde g_i=\left\{\tilde h_{r_i,d},\tilde
h_{r_i,e}\right\}$. Thus, according to (\ref{eq7}), the secrecy rate of
relay $i$ depends on $\tilde g_i$. The source node will choose $K$
relay nodes for transmitting according to the relay's reported
information $\hat g$. The principle of source node is to find the $K$
relays to maximize the secrecy rates. The outcome function can be
stated as
\begin{align}
F({\bf{\hat g}}) = \arg \max \sum\limits_{{R_i} \in \mathcal
{K}}^K {{C_{i,s}}({{\hat g}_i})}.
\end{align}
We define $\pi$ as the price per unit of secrecy rate achieved by
the relay. The relays in the network are assumed to be rational and
fair-minded, which means that although they are selfish, none is
malicious. The object of relay is to make itself chosen for
transmitting so that it can gain payoff. Due to the channel
orthogonality, the \emph{utility} of $R_i$ can be expressed as
\begin{align}
{D_i} = \left\{ {\begin{array}{*{20}{c}}
   {\pi {C_{i,s}},\;\;\;\;\;\;\;\;\;\;\;\;{R_i} \in {\cal K}},  \\
   {0,\;\;\;\;\;\;\;\;\;\;\;\;\;\;\mbox{otherwise}.}  \\
\end{array}} \right.
\end{align}
The total \emph{payoff (utility)} from the system can also be
expressed as
\begin{align}
D = \sum\limits_{i = 1}^K {{D_i}}.
\end{align}

We assume that the channel information is the private information of
each relay, and thus, the source is unable to know whether the
reported information is true or not. Since only the relay nodes
selected by the source for secure data transmission can get the
payoff, they will not report their true information to the source in
order to win greater opportunity to be selected. In this situation,
it can cause unfairness in selection and damage the expected payoff
of those unselected. It can also decrease the total payoff paid by the
system which can be expressed as
$\hat D \leq \tilde D,$
where $\hat D$ represents the total the total payoff
calculated according to the information reported by the relay nodes. $\tilde D$ represents
the total payoff when all the relay nodes report the truth. These results can sabotage the reliability of the system
and eavesdropper can easily sniff the transmitted messages.

Firstly, we prove that no equilibrium can be achieved under this
condition.

\textbf{Proposition 1}: Assuming that $R_i$ does not know other
relay's channel information, respectively, secrecy rate. But it
knows that each relay obeys a certain probability density function
defined as $p\left({{\tilde C}_{j,s}}\right) \left(0 \le {{\tilde C}_{j,s}} < \infty
,j \ne i\right)$. Then, $R_i$ has an incentive tendency to exaggerate its ${{\hat
C}_{i,s}}$ to $\infty$ to get the maximum expected payoff.

\begin{proof} $R_i$'s expected payoff can be also be expressed as
\begin{align}\label{eq18}
{D_i}({{\hat g}_i}) = \pi {{\tilde C}_{i,s}}{\rm{P}} ({R_i} \in
\mathcal {K}),
\end{align}
where $P\left(R_i \in \mathcal{K}\right)$ represents the probability of $R_i$
when being chosen. Considering the principle of choosing relay,
$P\left(R_i \in \mathcal{K}\right) \propto {\hat C_{i,s}}$ and when $\hat
C_{i,s}\to \infty$, $P\left(R_i \in \mathcal{K}\right)\to1$, so that $R_i$ gets
its maximum payoff at infinity. This indicates that every relay node
has the incentive to exaggerate its channel information to the
source, and thus, there is no equilibrium achieved under this kind
of payoff allocation.
\end{proof}

\subsection{VCG-based Mechanism Solution }
In order to prevent relay nodes from reporting distorted channel information, we propose an effective self-enforcing truth-telling mechanism to solve this problem. By using the VCG-based mechanism, the honest relay nodes gain the maximum payoff, as any cheating in the process will lead to decrease in payoff. Like the VCG mechanism, we introduced this transfer payment of  $R_i$  as
\begin{align}  \label{eq26}
{t_i}\left({{\hat g}_i},{{\hat g}_{ - i}}\right)=\sum\limits_{j \ne i}^{K}
{{D_j}({{\hat g}_j})}  - \sum\limits_{j \ne i}^{K}
{D_j^*({{\hat g}_j})},
\end{align}
where $D_j^*(.)$ denotes the utility of $R_j$  when $R_i$
does not participate in the system. So the total payoff of $R_i$ is:
\begin{align}\label{eq20}\nonumber
{U_i}\left({{\hat g}_i}\right) &= {D_i}({{\hat g}_i}) + {t_i}({{\hat
g}_i},{{\hat g}_{ - i}}) \\ &= {D_i}({{\hat g}_i}) + \sum\limits_{j
\ne i}^{N} {{D_j}({{\hat g}_j})}  - \sum\limits_{j \ne
i}^{K} {D_j^*({{\hat g}_j})}
\\& = \sum\limits_{j}^K {{D_j}({{\hat
g}_j})} - \sum\limits_{j \ne i}^{K} {D_j^*({{\hat
g}_j}).}\nonumber
\end{align}
If one relay node claims a higher $\hat h_{r_i, d}$ or a lower $\hat h_{r_i, e}$ than the reality to make its secrecy rate larger, it may get more chances to be selected by the source node, but also will pay a higher transfer payoff to those unselected. On the contrary, if one relay node reports a lower secrecy rate than reality, it will receive the compensation from other relay nodes at the cost of less chances to be selected. By adding this transfer function, we will discuss some properties of this VCG-based mechanism as follows.

\textbf{Proposition 2}: By using the VCG transfer function (\ref{eq26}) to balance the payoff allocation, relay node $R_i$ can gain its largest payoff when it reports the true private channel information.

\begin{proof} We can see from (\ref{eq20}) that the payoff of each
relay $R_i$  is the total utility of all relays $\sum_j^K
{{D_j}({{\hat g}_j})}$ when relay participates in the system, minuses
the total utility of all other relays $\sum_{j \ne
i}^{K} {D_j^*({{\hat g}_j})}$ when relay $i$ withdraws from the
system. It is obvious that relay $i$ cannot influence the value of
$\sum_{j \ne i}^{K} {D_j^*({{\hat g}_j})}$. Therefore, in
order to maximize its own payoff, relay $i$ seeks to maximize
the total utility of the system. According to our relay selection
principle, the total utility of all relays depends on the chosen $K$
relay's true channel information. If and only if each relay reports
the true information $(\hat g_i = \tilde g_i)$, the total utility is
maximized. Hence, the payoff of $R_i$  is maximized.
\end{proof}
\textbf{Proposition 3}: Every rational relay node in the system takes part in the VCG-based mechanism for its own benefit.

\begin{proof} It is easy to show that $\sum_j^K {{D_j}({{\hat
g}_j})} \ge \sum_{j \ne i}^{K} {D_j^*({{\hat g}_j})}$
when the IC achieved $(\hat g_i = \tilde g_i)$, and the equality holds when
$R_i$ is not selected ($D_i = 0$). Therefore, for each relay $R_i$ ,
$U_i(g_i) \ge 0$ and participating into the system is an optimal
choice for a rational relay.
\end{proof}
\textbf{Proposition 4}: By applying the VCG-based mechanism in our system, we cannot achieve the BB condition: the total transfer payments $\sum_{i =
1}^N {{t_i}}  < 0$, which means that we need the mechanism designer or a third party to pay parts of the payoff.
\begin{proof} There are two cases of $R_i$:
\begin{itemize}
\item It is not selected by the source node (${R_i} \notin \mathcal
{K}$), then obviously: ${t_i}\left({{\hat g}_i},{{\hat g}_{ - i}}\right) =
\sum_{j \ne i}^{K} {{D_j}({{\hat g}_j})}  -
\sum_{j \ne i}^{K} {D_j^*({{\hat g}_j})} = 0$.

\item It is selected by the source node $(R_i \in \mathcal{K})$, then
${t_i}({{\hat g}_i},{{\hat g}_{ - i}}) = \sum_{j \ne
i}^{K} {{D_j}({{\hat g}_j})}  - \sum_{j \ne i}^{K}
{D_j^*({{\hat g}_j})} < 0$. Because of the withdrawal of $R_i$,
another relay node with a lower secrecy rate will be selected and
its utility $D^*$ will get bigger.
\end{itemize}

Combine these two cases together: $K$ relay nodes will receive
negative transfer payment that makes the total transfer payments
$\sum_{i = 1}^K {{t_i}}  < 0$. Hence, the BB is not satisfied
in the VCG-based mechanism.
\end{proof}

\subsection{AGV-based Mechanism Solution}
From the discussions above we can know that the VCG-based mechanism can enforce every relay node to tell the true private channel information, which can effectively solve the cheating problem in our system. However, as the mechanism designer, we need to pay some extra payments to the system because the VCG-based mechanism fails the  condition of BB. To compensate for this loss, we improve the VCG-based mechanism to the AGV-based one.

In the AGV-based mechanism, we change the transfer payment of $R_i$ as
\begin{align}\label {eq21}
{t_i}({{\hat g}_i},{{\hat g}_{ - i}}) = {\Phi _i}({{\hat g}_i}) -
\frac{1}{{K - 1}}\sum\limits_{j \ne i}^K {{\Phi _j}({{\hat g}_j})}
\end{align}
where
\begin{align}\label{eq22} \nonumber
{\Phi _i}\left({{\hat g}_i}\right) &= {E_{{\hat g_{ - i}}}}\left[\sum\limits_{j = 1,j
\ne i}^K{D_j}({{\hat g}_j})\right] \\
&= \sum\limits_{j = 1,j \ne i}^K
{{E_{{\hat g_{ - i}}}}\left[{D_j}({{\hat g}_j})\right]}
\end{align}
represents the sum of the other relay nodes' expected utilities
given the reported information $\hat g_i$.

Like in the VCG-based mechanism, we can prove that only if the relay nodes reveal the true channel information, they can obtain the maximum payoff in the AGV-based mechanism. There is only one equilibrium under this kind of payoff allocation.

\textbf{Proposition 5}: By using the AGV-based mechanism, the relay node
$R_i$ can gain its largest expected payoff when it reports its true
private channel information to the source node.

\begin{proof} Without loss of generality, we consider the expected
payoff of $R_1$. Since $R_1$ only knows its own channel information,
we can calculate the payoff according to the transfer payment
function (\ref{eq21}) as

\begin{align}\label{eq23}\nonumber
 &E[{U_1}({{\hat g}_1})] = E[{D_1}({{\hat g}_1}) + {t_1}({{\hat g}_1},{{\hat g}_{ - 1}})] \\ \nonumber
  &= E[{D_1}({{\hat g}_1})] + {E_{{{\hat g}_{ - 1}}}}\hspace{-0.1cm}\left[\sum\limits_{j \ne 1}^K {D_j}({{\hat g}_1})\right]\hspace{-0.1cm}  - \hspace{-0.1cm}\frac{1}{{K - 1}}\sum\limits_{j \ne 1}^K {{\Phi _j}({{\hat g}_j})}  \\
  &= E\left[\sum\limits_{j = 1}^K{D_j}({{\hat g}_j})\right] -  \frac{1}{{K - 1}}\sum\limits_{j \ne 1}^K {{\Phi _j}({{\hat g}_j})}.
\end{align}
We can see that there are two terms in the right side of
(\ref{eq23}). The first one represents the total expected payoff
when $R_1$ reports $\hat g_1$ as its channel information (the
expectation is calculated by $R_1$ itself). Since the other term
being independent of $\tilde g_1$, only the first term decides the expected
payoff of $R_1$. As we have shown above, the total payoff is based
on the real secrecy rate. Only when the $K$ relays with top $K$ secrecy rate
are selected, the total payoff will be maximized. Any cheating leads
to a decrease in all relays' total payoff, and therefore, the
expectation $E[U_1(\hat g_1)]$ can get the maximum when $R_1$
reports its true channel information. Similarly, each relay node in
the network has an incentive to report its true channel information
$({{\hat g}_i} = {{\tilde g}_i})$. Thus, the equilibrium is achieved
under this condition.
\end{proof}
\textbf{Proposition 6}: Each relay node could gain a positive
expected payoff in the AGV-based mechanism, which ensures that every relay would like to take part in this mechanism.

\begin{proof} From (\ref{eq21}) and (\ref{eq22}) it is easy to
derive $R_i$'s expected payoff:
\begin{align}\nonumber
 E[{U_i}({{\hat g}_i})] &= \hspace{-0.1cm}E\hspace{-0.1cm}\left[\sum\limits_{j = 1}^K {D({{\hat g}_j})} \right]\hspace{-0.1cm} - \hspace{-0.1cm}\frac{1}{{K - 1}}\sum\limits_{j \ne i}^K {\sum\limits_{k \ne j}^K {{E_{ - k}}[{D_k}({{\hat g}_j})]} }
 \\\nonumber
 & = \hspace{-0.1cm}E\hspace{-0.1cm}\left[\sum\limits_{j = 1}^K {D({{\hat g}_j})} \right]\hspace{-0.1cm} - \hspace{-0.1cm}\frac{1}{{K - 1}}\left\{ (K - 1)\sum\limits_{j = 1}^K {E[{D_j}({{\hat g}_j})]} \right.\\
 & - \left.\sum\limits_{k \ne i}^K {{E_{ - k}}[{D_k}({{\hat g}_i})]} \right\}
 \\\nonumber
 &= \frac{1}{{K - 1}}{E_{ - j}}\hspace{-0.1cm}\left[\sum\limits_{j \ne i}^K {{D_j}({{\hat g}_i})} \right].
\end{align}
According to (\ref{eq18}), $D_i>0$ if $R_i$ is selected and $D_i =
0$ if not. Since among all the relay nodes there are always some
nodes being selected, the right side of the equation above ${E_{ -
j}}\left[\sum\nolimits_{j \ne i}^N {{D_j}({{\hat g}_i})} \right] > 0$. Therefore,
$R_i$ can gain a payoff more than 0, and thus, the IR is satisfied.
\end{proof}

 \textbf{Proposition 7}: In the AGV-based mechanism, the system can achieve budget
balance, which means that we, as the mechanism designer, will not pay any extra payment to the system.

\begin{proof} If we calculate the total transfer payment of all
relays, we could get
\begin{align}\nonumber
 \sum\limits_{i = 1}^N {{t_i}({{\hat g}_i},{{\hat g}_{ - i}})}  &= \sum\limits_{i = 1}^N {{\Phi _i}({{\hat g}_i})}  - \frac{1}{{N - 1}}\sum\limits_{i = 1}^N {\sum\limits_{j = 1,j \ne i}^N {{\Phi _j}({{\hat g}_j})} }  \\
  &= \sum\limits_{i = 1}^N {{\Phi _i}({{\hat g}_i})}  - \sum\limits_{j = 1}^N {{\Phi _j}({{\hat g}_j})}  = 0.
\end{align}
This implies that the proposed transfer function can realize a
payment reallocation among the relay nodes, and no extra payment is
required to be paid by the system or by the relay nodes.
\end{proof}

\subsection{The Value of K}
In the discussion above, we assumed that the value of $K$ is fixed and
the source node always choose $K$ relays for cooperating. However, it
is easy to see that different $K$ can lead to different results in
the total secrecy rate of the network. So we did some research to
figure out the optimal amount $K$ of relays the source node should
select.

In our system model, the total secrecy attained of the system is
(\ref{eq15}). By using the AGV mechanism, each relay reports the
truth. Now we assume each relay's reported information is
$\mbox{SNR}_{r_i,d}$ and $\mbox{SNR}_{r_i,e}$. Let $k_i =
\frac{\mbox{SNR}_{r_i,d}}{\mbox{SNR}_{r_i,e}}$. Sort $k_i$ in
descending order, and get $k_{(1)}\ge k_{(2)} \ge \ldots \ge
k_{(N)}, k_{(i)} \in \{k_1, k_2, \ldots, k_N\}$. Obviously, the
relay which has a larger $C_{i,s}$ also has a larger $k_i$ according
to (\ref{eq7}). We denote that $R_{(1)}$ is the best relay which has
the largest secrecy rate, $R_{(2)}$ is the second best, and so forth.
Then the optimal selection strategy of the source node is described
as below:

1. Select $R_{(1)}$ for transmitting. Let $i= 1$ and calculate
$\Psi_1 = \frac{1+\mbox{SNR}_{s,d}+\mbox{SNR}_{r_{(1)},d}}
{1+\mbox{SNR}_{s,e}+\mbox{SNR}_{r_{(1)},e}}$.

2. For $i < N$, if $\Psi_i < k_{(i+1)}$, proceed step 3 and if
$\Psi_{i} \ge k_{(i+1)}$, skip to step 4.

3. Select $R_{(i+1)}$ and calculate ${\Psi _{i + 1}} = \frac{{1 +
{\rm{SN}}{{\rm{R}}_{s,d}} + \sum\nolimits_{j = 1}^{i + 1}
{{\rm{SN}}{{\rm{R}}_{{r_{(j)}},d}}} }}{{1 +
{\rm{SN}}{{\rm{R}}_{s,e}} + \sum\nolimits_{j = 1}^{i + 1}
{{\rm{SN}}{{\rm{R}}_{{r_{(j)}},e}}} }}$. Then let $i = i+1$ and go
back to step 2.

4. Let $K$ = i and stop.

\textbf{Proposition 8}: The system can attain the largest secrecy
rate by selecting $K$ relays for transmitting data, where $K$ is
decided by the process above.

\begin{proof} By the selection strategy of the source described
above, it is easy to prove that $\Psi_{K}$ is the maximum among
$\left\{\Psi_1, \Psi_2, \ldots, \Psi_{N}\right\}$. According to (\ref{eq15}), the
total secrecy rate of the network when selecting $i$ relays can be
expressed as $C_{s,sys}(i)=W\log_2\Psi_i$. When $i=K$, $\Psi_{K}$ can
get the maximum, and obviously $C_{s,sys}(K)$ is the largest.
Therefore, it is the best choice for the source to select $K$ relays
in the system.
\end{proof}

In many cases, because of the geographic conditions, the direct
transmission is very weak compared with relay transmission which
means that the selected relay has a $\mbox{SNR}_{r_i,d}
> \mbox{SNR}_{r_i,e} \gg \mbox{SNR}_{s,d} > \mbox{SNR}_{s,e}$. So $\mbox{SNR}_{r_i, d} \gg (1+\mbox{SNR}_{s,d}), \mbox{SNR}_{r_i,
e} \gg (1+\mbox{SNR}_{s,e})$ and $\Psi_1 \approx k_{(1)}
> k_{(2)}$. Thus, $K=1$ is the best choice, which means the source
should only select one best relay for transmitting data. More than
one relay node would lead to a decrease in the total secrecy rate of
the system.

\section{Simulation Results}%

In this section, we provide simulation results of the wireless relay
system in the VCG-based mechanism and AGV-based mechanism, respectively.
Specifically, to simplify the calculation and simulation, we assume
that each relay node first calculates its own secrecy rate according
to its channel information, and then reports it to the source.
Without considering the process of calculating $\pi C_{i,s}$, we
assign random values $x_i$ to indicate $\pi C_{i,s}
(i=1,2,\ldots,N)$, which not affect the ``outcome" or source's
selection result. Furthermore, we assume that though $R_i$ does not
know other relays' channel information, it knows that every reported
value obeys the probability density function: $e^{ - x_i }$ $\left(x_i
\in [0,\infty ) \;\mbox{and}\; \int_0^{ + \infty } {e^{ - x_i } dx_i }
= 1\right)$.

Firstly, we consider a system with $N=4$ relay nodes and from which
the source node chooses $K=2$ relays. A random sample of these relay
nodes' secrecy rates is obtained as ${[1.0132, 0.6091, 0.3885,
1.3210]}$ and the price per unit of secrecy rate $\pi = 1$ is
assumed.

\begin{figure}[!t]
\centering
\includegraphics[width=4.8in]{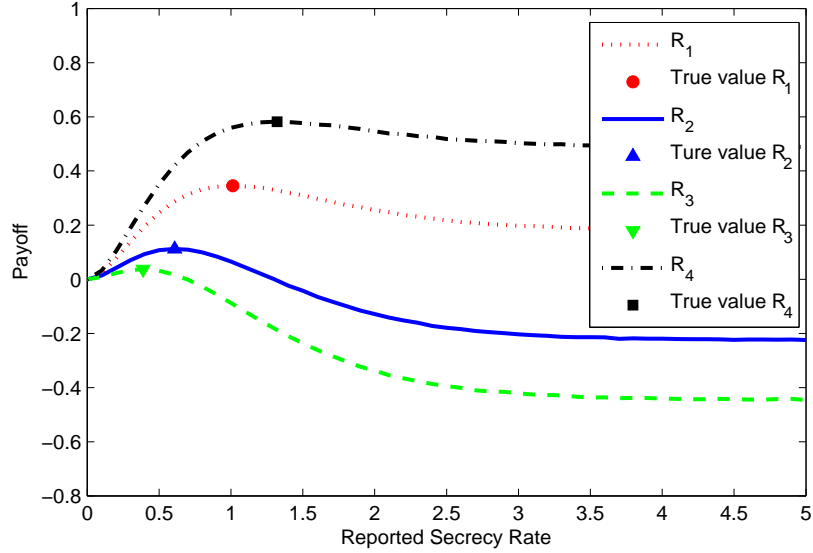}
\caption{{Payoff of $R_i$ when different secrecy rates are reported
in the VCG-based mechanism.}} \label{fig_2}
\end{figure}

Fig.~\ref{fig_2} shows the variation of $R_i$'s payoff when the
reported values change in the VCG-based mechanism. Given that the other three
nodes are honest, $R_i (i=1,2,3,4)$ can get its maximum payoff
while reporting the truth. From Fig.~\ref{fig_2} we can observe that
when they all tell the truth, the larger the true value of secrecy
rate of one relay node is, the more the payoff it gains. For
example, $R_4$ has the largest secrecy rate ($ \tilde C_{4,s} =
1.3210$) and its payoff is the largest up to $0.5822$ when it
reports the true value. It is higher than the other three relay
nodes' payoff even though it is not as much as $\pi C_{4,s} =
1.3210$, which is paid by the destination node because of the
transfer payment.

\begin{figure}[!t]
\centering
\includegraphics[width=4.8in]{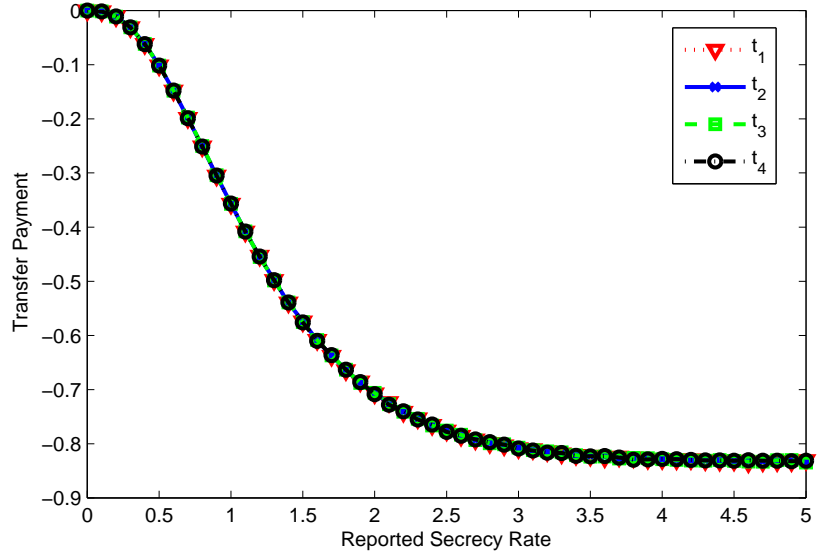}
\caption{Transfer payment of $R_1$ when different secrecy rates are
reported in the VCG-based mechanism.} \label{fig_3}
\end{figure}

In Fig.~\ref{fig_3} we demonstrate the transfer payment of each
relay they calculate from their own angles when they report
different secrecy rates. As is evident in the figure, each relay has
the same transfer payment curve. This is because we assume each
relay only knows the other relays report secrecy rates obey the
negative exponential distribution. So the difference of the utility
of the others relays whether the relay participates the mechanism or
not is the same for each relay. We can also see that they are all
monotone decreasing because the larger the reported value is, the
more transfer payoff should be paid to others. Besides, as the reported secrecy rate continuously increases, the transfer payoff will tend to a fixed value. It is because this very large reported value will always be larger than the others, the ``outcome" or source's selection will be fixed whether this relay node is in this system or not. So the transfer payment will be a fixed value when the reported value becomes very large. The curve of $R_i$'s payoff in Fig.~\ref{fig_2} is the same reason.

\begin{figure}[!t]
\centering
\includegraphics [width=4.8in]{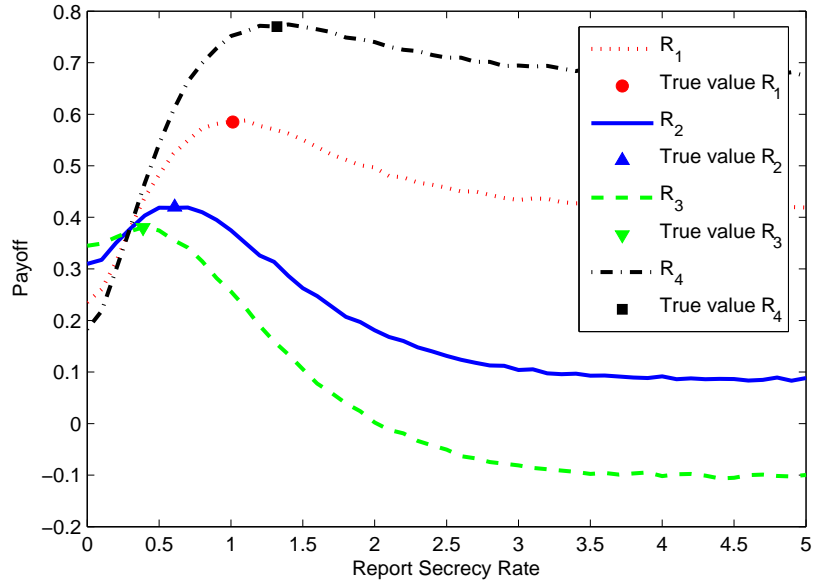}
\caption{Expected payoff of $R_i$ when different secrecy rates are
reported in the AGV-based mechanism.} \label{fig_4}
\end{figure}

\begin{figure}[!t]
\centering
\includegraphics[width=4.8in]{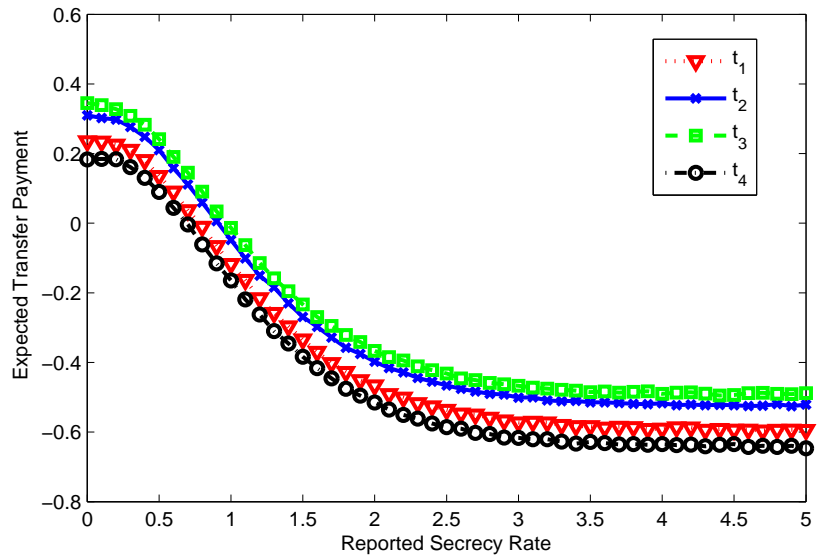}
\caption{Expected transfer payment of $R_i$ when different secrecy
rates are reported in the AGV-based mechanism.} \label{fig_5}
\end{figure}

\begin{figure}[!t]
\centering
\includegraphics[width=4.8in]{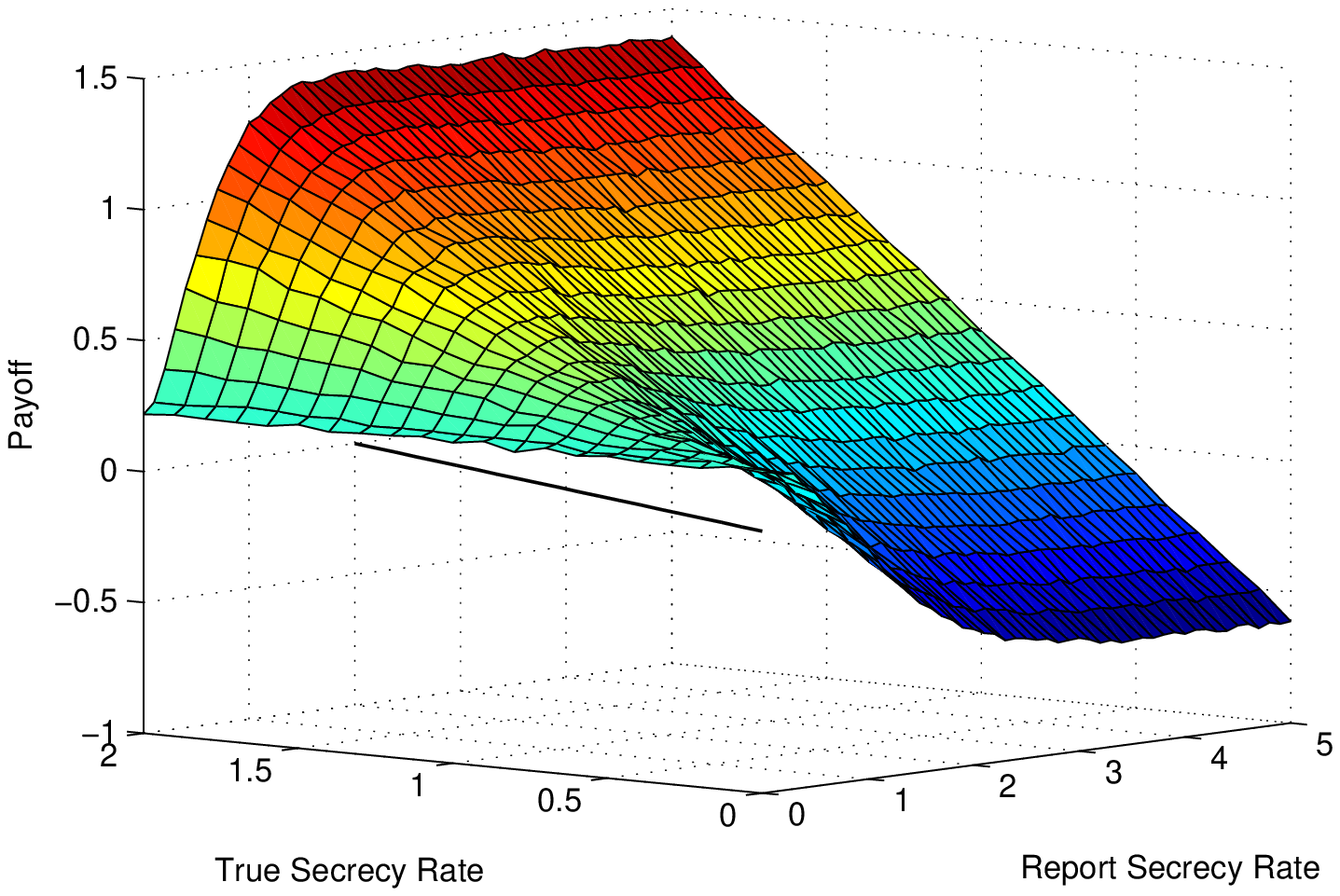}
\caption{Expected payoff of $R_1$ with variable true secrecy rate when different secrecy
rates are reported in the AGV-based mechanism.} \label{fig_3d}
\end{figure}

Fig.~\ref{fig_4} and Fig.~\ref{fig_5} show the results when we use
the AGV-based mechanism in the system. In Fig.~\ref{fig_4}, four curves
show the expected payoff of $R_1$-$R_4$. It is obvious to see that
each relay maximizes its payoff when they report the true secrecy
rate. Compared with Fig.~\ref{fig_2}, we can see each relay's payoff
is higher in the AGV-based mechanism than that in the VCG-based mechanism. It means that the
AGV-based mechanism can maximize all the relay nodes payoff which is
more attractive for relay node to attend. From Fig.~\ref{fig_5} we
also find that $R_1$'s and $R_4$'s transfer payoffs are negative
while the other two's are positive when they tell the truth. This is
because $R_1$ and $R_4$ are actually selected by the source node and
need to pay the transfer payment while $R_2$ and $R_3$ are not. By
using the AGV-based mechanism, the relay nodes with smaller secrecy rate
will get compensations from those with larger ones. It can balance
the payment allocation of the system and benefit those in worse
physical conditions. Furthermore, we calculate the expected transfer
payoff of $R_i$ when they all report the truth: $t_1=-0.1247$,
$t_2=0.1570$, $t_3=0.2831$, $t_4=-0.3154$ and $t_1+t_2+t_3+t_4=0$,
which is in accord with the equation (\ref{eq18}).
Hence the system is budget balanced and no extra payment is paid
into or out of the network. In conclusion, we can confirm that the AGV-based
mechanism is more compatible for our system than the VCG-based mechanism.
Moreover, we show the payoff of $R_1$ with changing secrecy rate when it reports different secrecy rate
in Fig.~\ref{fig_3d}. It is obvious that no matter what true secrecy rate of $R_1$ is, $R_1$ always gains its maximum payoff when it reports its own true secrecy rate.

\begin{figure}[!t]
\centering
\includegraphics[width=4.8in]{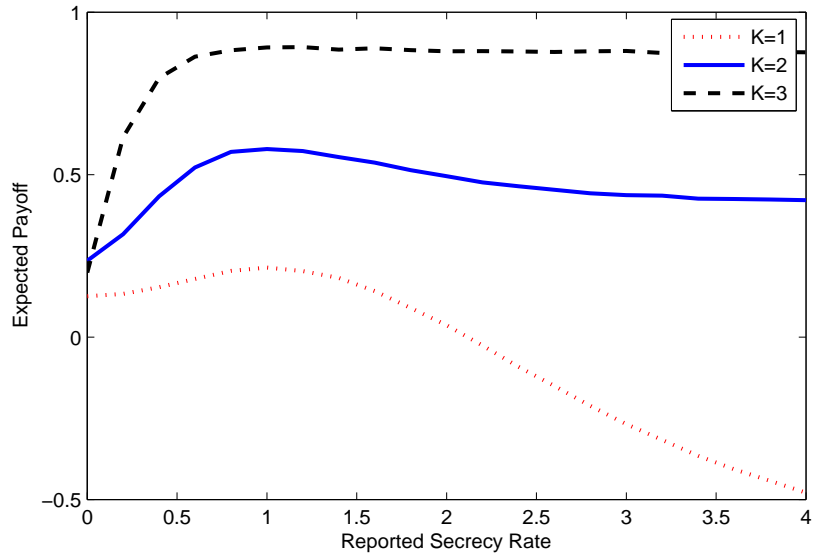}
\caption{Effectiveness of the reported secrecy of $R_1$ on the expected payoff at different $K$ in the AGV-based
mechanism.} \label{fig_6}
\end{figure}

\begin{figure}[!t]
\centering
\includegraphics[width=4.8in]{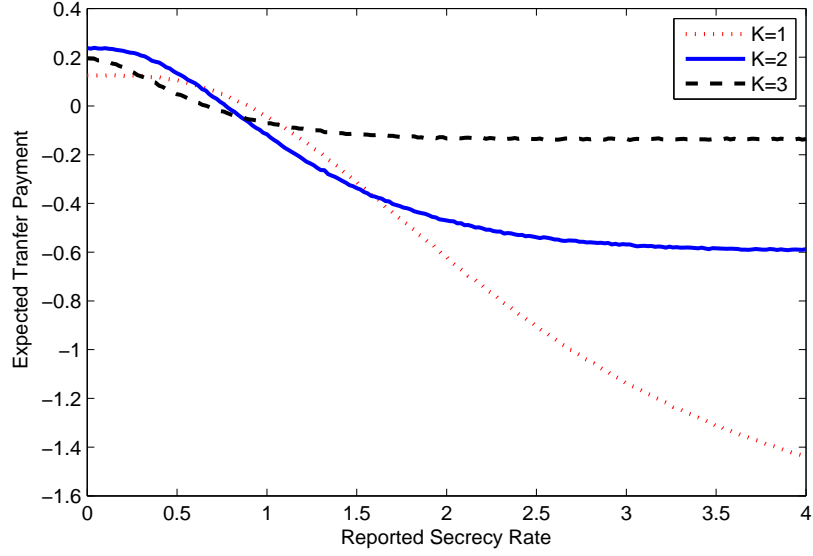}
\caption{Effectiveness of the reported secrecy of $R_1$ on the expected transfer payment at different $K$ in the AGV-based
mechanism.} \label{fig_7}
\end{figure}

In addition, we analyze the effects of the reported
secrecy rate versus the value of $K$ in the AGV-based mechanism. As an example, we set the relay node $R_1$ be
the interested one and its secrecy rate is $1.0132$. In Fig.~\ref{fig_6}, we observe that the relay node $R_1$ achieve the
maximum expected payoffs when it reports the true secrecy rate with different $K$. When $K$ equals to $1$, the payoff of the relay node $R_1$ is the lowest. And the larger the reported value, the smaller the expected payoff shows. When $K$ equals $2$ and $3$,
the expected payoff becomes a fixed value as the reported secrecy rate continuously increases.
Because the transfer payment is a part of the total payoff,
the change of the expected payoff can be translated by the transfer payment, which is shown in Fig.~\ref{fig_7}. When $K$ equals to $1$
and this relay reports a larger secrecy rate, the expected transfer payment is a small and negative value. So the expected payoff
of the relay node is minor. When $K$ equals to $2$ and $3$, the slope of the curves becomes smoother as the reported value increases.
This shows the change trend of the relay node's payoff from the other point of view. Meanwhile, it implies that the transfer payoff is
helpful to control the payoff for fairness among relay nodes.

\begin{figure}[!t]
\centering
\includegraphics[width=4.8in]{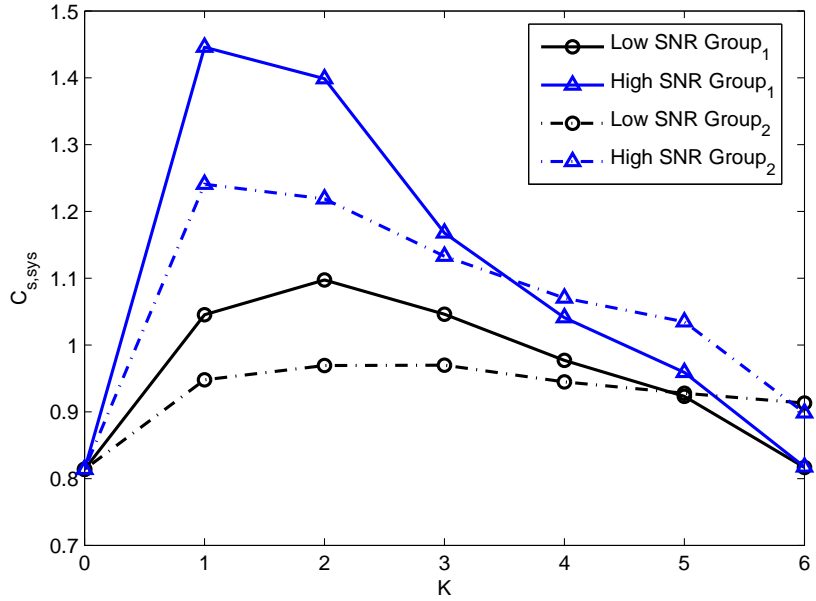}
\caption{System secrecy rate at different $K$ at bad SNR in the AGV-based
mechanism.} \label{fig_8}
\end{figure}

Finally, we focus on the effect of the value of $K$ on the total system secrecy.
 Here we assume $W=\ln2$ then
$C_{s,sys}(K)=\ln(\Psi(k))$. Let $N=6$, the direct transmission
$\mbox{SNR}$ to destination and eavesdropper are $9.64$dB and
$5.47$dB, respectively, and given two random samples for $R_i$'s
report information ($\mbox{SNR}_{r_i,d}$ and $\mbox{SNR}_{r_i,e}$).
In one sample we assume each value has the same order of magnitude
with
$1$ $(1\mbox{dB}<\mbox{SNR}<10\mbox{dB}):\mbox{SNR}_{r_i,d}=\{6.1734,$ $7.9489,$ $9.7429,$ $7.1886,$ $6.3783,$ $7.3411\}, (\mbox{dB})$;
$\mbox{SNR}_{r_i,e}=\{3.7700,$ $0.9927,$ $5.6543,$ $4.3645,$ $0.6273,$ $6.1954\}, (\mbox{dB})$. In the other sample we assume each value has the
same order of magnitude with
$10$ $(10\mbox{dB}<\mbox{SNR}<20\mbox{dB}):\mbox{SNR}_{r_i,d}=\{16.173,$ $17.948,$ $19.742,$ $17.188,$ $16.378,$ $17.341\}, (\mbox{dB});
$ $\mbox{SNR}_{r_i,e}=\{13.770,$ $10.992,$ $15.654,$ $14.364,$ $10.627,$ $16.1954\}, (\mbox{dB})$. Similarly, we do this simulation with another group data of high SNR and low SNR as follows: $\mbox{SNR}_{r_i,d}=\{8.8149,$ $5.6809,$ $9.3701,$ $8.5822,$ $3.3896,$ $10.000\}, (\mbox{dB})$; $\mbox{SNR}_{r_i,e}$ $=\{3.7700,$ $0.9927,$ $ 5.6543,$ $4.3645, $ $0.6273,$ $6.1954\}, (\mbox{dB})$; $\mbox{SNR}_{r_i,d}$ $=\{$$16.173,$ $17.948,$ $19.742,$ $17.188,$ $16.378,$
$17.341\}, (\mbox{dB})$; and $\mbox{SNR}_{r_i,e}=\{$$15.227,$ $12.164,$ $14.522,$ $13.278,$ $12.746,$ $13.648\}, (\mbox{dB})$. The simulation result is showed in Fig.~\ref{fig_8}, and we can
observe that in the low $\mbox{SNR}$ situation, $C_{s,sys}$ is maximized when the
source select 2 and 3 relays, respectively. Thus, they attain the maximum secrecy at
$K=2$ and $K=3$. However, in the high $\mbox{SNR}$ situation showed in Fig.~\ref{fig_8},
when all of the channel conditions are better, the best choice for
the system is to choose only one relay $(K=1)$ for transmitting. All
these results are based on the fact that all relays will reveal
their true channel information which is ensured by the AGV mechanism.

\section{Conclusions}%

In this paper, we discussed and applied the ideas of mechanism
design into the wireless relay network to guarantee the
strategy-proof during the process of relay selection when considering secure
data transmission. We proved that by using the VCG
mechanism and AGV mechanism, each relay node gets its maximum payoff
only when it reveals its true channel information, and any deviation
from the truth will lead to a loss in its own (expect) payoff as
well as the total secrecy rate. We compared these two mechanisms and
illustrated that the AGV mechanism is more compatible for our system
when taking the budget balance constraint into consideration. We
proved that the strategy-proof and budget balance of the system
can be achieved in the AGV mechanism, which makes our model more
practical in reality. Simulation results verified these conclusions.
Moreover, we proposed and proved the best choice for the source node
in deciding how many relays it should select to get the maximum
secrecy rate of the network. In good channel conditions with higher
$\mbox{SNR}$, it is better to select only one relay for transmitting
data.

\appendices
\end{document}